\documentclass{aa}
\pdfoutput=1
\bibpunct{(}{)}{;}{a}{}{,}
\usepackage{natbib}
\usepackage[varg]{txfonts}
\usepackage{graphicx}

\begin{document}

\title{Spectroscopic observation of a transition region network jet}

\author{J. Gorman\inst{1}
\and L.P. Chitta\inst{1}
\and H. Peter\inst{1}}

\institute{Max Planck Institute for Solar System Research, Justus-von-Liebig-Weg 3, 37077 Göttingen, Germany}

\date{Received DD Month YYYY / Accepted DD Month YYYY}

\abstract {} {Ubiquitous transition region (TR) network jets are considered to be substantial sources of mass and energy to the corona and solar wind. We conduct a case study of a network jet to better understand the nature of mass flows along its length and the energetics involved in its launch.} {We present an observation of a jet with the Interface Region Imaging Spectrograph (IRIS), while also using data from the Solar Dynamics Observatory (SDO) to provide further context. The jet was located within a coronal hole close to the disk center.} {We find that a blueshifted secondary component of TR emission is associated with the jet and is persistent along its spire. This component exhibits upward speeds of approximately 20--70~km\,s$^{-1}$ and shows enhanced line broadening. However, plasma associated with the jet in the upper chromosphere shows downflows of 5--10~km\,s$^{-1}$. Finally, the jet emanates from a seemingly unipolar magnetic footpoint.} {While a definitive magnetic driver is not discernible for this event, we infer that the energy driving the network jet is deposited at the top of the chromosphere, indicating that TR network jets are driven from the mid-atmospheric layers of the Sun. The energy flux associated with the line broadening indicates that the jet could be powered all the way into the solar wind.}

\keywords{Sun: atmosphere -- Sun: chromosphere -- Sun: transition region -- Line: profiles}

\maketitle

\section{Introduction}
\label{section:Introduction}
The study of the transfer of energy from the cool, lower solar atmosphere through the transition region (TR) in order to sustain the hot corona remains an active field of research within solar physics. In particular, localized energetic events such as spicules and jets are believed to be important contributors of mass and energy to both coronal heating and the solar wind. Jets are seen at a broad range of temperatures and spatial scales, from hotter ($\approx$\,1~MK) coronal jets with broader spires ($\approx$\,10~Mm) \citep[reviewed in e.g.,][]{raouafi:2016} to much cooler ($10^4$~K) and narrower (down to a few hundred kilometers) chromospheric anemone jets \citep{nishizuka:2011}.

Recent observational works have been able to explore the ubiquity of jets in the TR using the Interface Region Imaging Spectrograph \citep[IRIS;][]{depontieu:2014_07}. Transition region network jets are seen as collimated and propagating intensity fronts. These jets are ubiquitous and are generally rooted to the boundaries of network magnetic patches in the photosphere. They were first identified by \citet{tian:2014} with near-limb observations using the 1330~\AA\ channel of IRIS' slit-jaw imager (SJI). These jet events were found to have characteristic velocities of 80--250~km\,s$^{-1}$, widths up to 300~km, and lifetimes of 20--80~s, reaching temperatures up to $10^5$~K. A similar study done by \citet{narang:2016} compared network jets in the quiet Sun (QS) to their counterparts in coronal holes (CHs). The authors found that jets in CHs were longer and faster than QS jets. These differences were due to the effect of (interchange) reconnection with open field lines in CHs compared to that with the smaller, closed loops found in the QS.

Various observational and simulation studies have been designed to better understand the primary driving mechanisms behind small-scale jets, often pointing to the involvement of magnetic reconnection at different heights above the solar surface and/or upward moving shocks. \citet{panesar:2016,panesar:2018} and \citet{mcglasson:2019} have recently reported on the magnetic driver of on-disk QS coronal jets. They determined that flux cancelation at the neutral line of mini-filaments led to the eruption of these mini-filaments. This, in turn, resulted in the ejection of the jets as seen in TR plasma. A recent observational study conducted by \citet{qi:2019} examined the link between network jets and coronal plumes, concluding that jets are more energetic in regions with visible plumes. This led them to surmise that the stronger magnetic convergence witnessed in the regions with plumes creates an environment with faster shocks or more small-scale reconnection that would drive the more dynamic jets. From a simulation standpoint, \citet{yang:2018} show that flux emergence can trigger both warm ($\approx$\,$10^5$~K) network jets and cool ($\approx$\,$10^4$~K) spicules through reconnection and pressure gradients, respectively.

While the presence of TR jets has been well-established, a discussion regarding  the veracity of their high-speed mass flows remains open. Previous studies \citep[such as e.g.,][]{tian:2014,narang:2016,chen:2019} relied on narrow-band imagery and sit-and-stare spectroscopic scans, which made it difficult to do more than track intensity movements along the jets or transverse jet motions. There is a debate about whether or not the assumed shooting upward movement of plasma is mostly just propagating thermal fronts, wave motions, or some weighted combination of the aforementioned \citep{chintzoglou:2018,depontieu:2017}. Specifically, \citet{rouppe:2015} point to the need for further study of the individual line profiles to better understand noted blueward asymmetries in jet spectra. In this context, capturing an ongoing network jet event while it is happening, along with its spectroscopic properties from its photospheric footpoints into the TR, would shed light on the nature of driving mechanisms and potential mass motions. However, such a clear spectroscopic observation of a jet event is rarely captured.

We have investigated a network jet in both narrow-band imagery and IRIS spectroscopic rasters, tracing it from its signatures in the TR down through the chromosphere to its magnetic footpoint in the photosphere. Due to the advantageous timing and spatial coverage of the raster slit, we were able to follow the spectral evolution of the jet along its (projected) length. Here, we report on the results of this analysis and provide insights into the nature of the flow of mass and energy within small-scale jets.


\begin{figure}
\includegraphics[width=8.8cm]{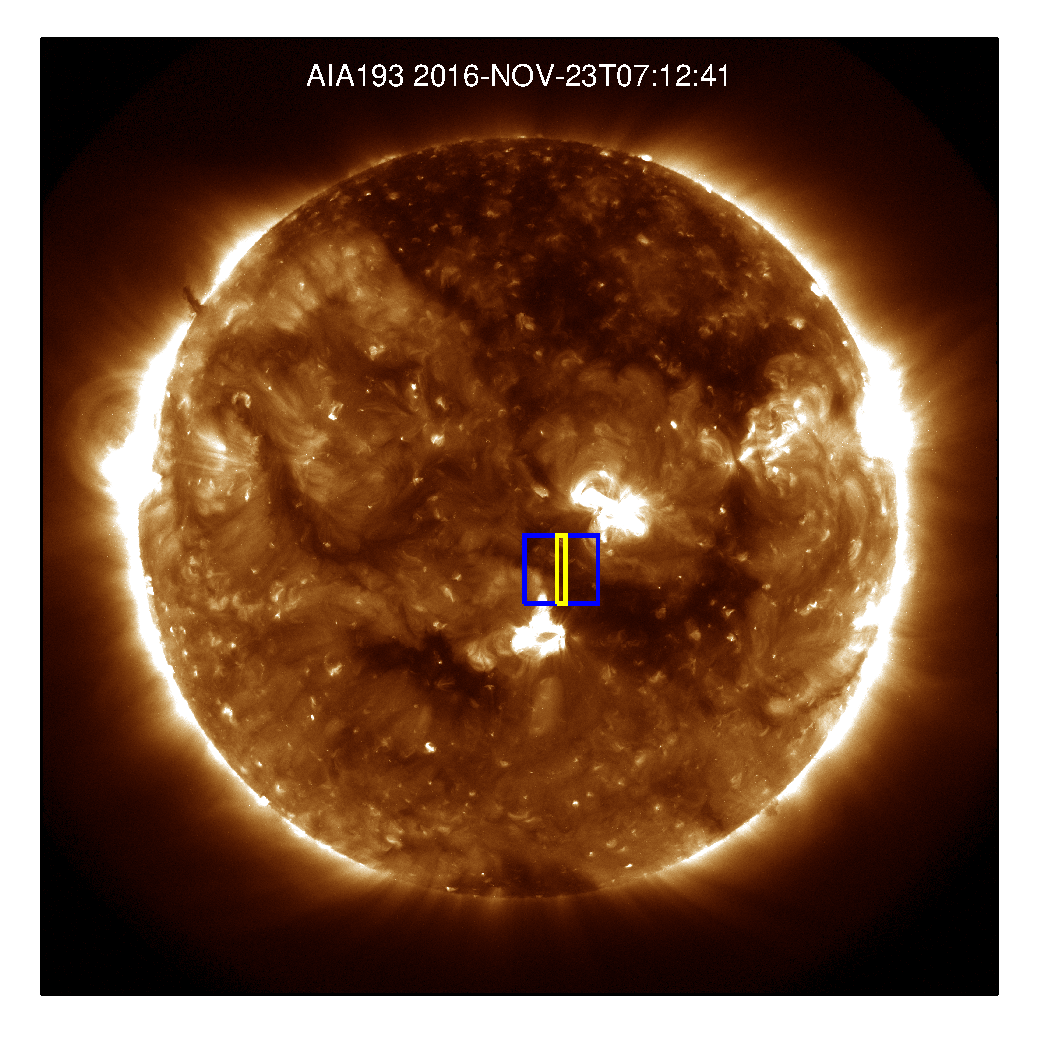}
\caption{Context image. AIA 193~\AA\ full-disk overview of the Sun on November 23, 2016, at 07:12:41~UT. The fields of view (FOV) for the IRIS raster and slit-jaw imagery are outlined in yellow and blue, respectively. Solar north is pointed up. See Sect.~\ref{section:Results}.}
\label{fig:AIA193overview}
\end{figure}
\section{Observations and data analysis}
\label{section:Observation}
IRIS observed an equatorial CH at disk-center bounded by two active regions (ARs) on November 23, 2016, from 07:12:43--07:46:33 UT. This observation, which is centered at $[x,\,y]=[110\arcsec,\,-134\arcsec]$, is comprised of a very-large dense raster scan with accompanying contextual slit-jaw imagery. The raster scanned the CH in 64 steps with 30~s exposure times (the step cadence is about 32~s, and the step size is 0.35\arcsec\ in the scan direction; the observation is spatially binned such that the image scale along the slit is 0.33\arcsec~pixel$^{-1}$). The raster covered a field of view (FOV) of $22\arcsec\times 175\arcsec$. The entire far-UV (FUV) and near-UV (NUV) detector windows were read out. All four of the slit-jaw channels (i.e., 1330, 1400, 2796, and 2832) were operating. For the purposes of this study, only the 1400 and 2796 SJI channels are discussed. These channels recorded 16 images each, at a cadence of 127~s. The total SJI FOV was $167\arcsec\times 175\arcsec$. For the SJI, similar to the raster, the data are binned in the y-direction. We also binned the SJI data in the x-direction during processing, resulting in a final pixel-scale of $[x,\,y]=[0.33\arcsec,\,0.33\arcsec]$. Within each IRIS SJI channel, we normalized the images by exposure time and aligned them to one another.

We complemented IRIS data with extreme ultraviolet (EUV) observations from the Atmospheric Imaging Assembly \citep[AIA;][]{lemen:2012} on board the Solar Dynamics Observatory \citep[SDO;][]{pesnell:2012}. In particular, we used AIA 193~\AA\ EUV filter images for context. These data have a cadence of 12~s and an image scale of 0.6\arcsec~pixel$^{-1}$. They were normalized by exposure time and co-registered to an AIA reference map taken at 07:27:54 UT on the day of the observation.

To study the magnetic field distribution underlying IRIS' FOV, we used line of sight (LOS) magnetic field maps from the Helioseismic and Magnetic Imager \citep[HMI;][]{scherrer:2012} on board SDO. These HMI data have a cadence of 45~s, an image scale of about 0.5\arcsec~pixel$^{-1}$, and were co-registered to the reference AIA image.

Using a cross-correlation technique, we aligned the SDO and IRIS data to enable easier feature location and comparison. The final product shown here is an aligned set of AIA, HMI, and IRIS SJI maps that share the same pixel size $[0.33\arcsec,\,0.33\arcsec]$ and orientation, with solar north pointed up.

In addition to the co-aligned imagery, the original, high-resolution spectra acquired by the IRIS raster scan were analyzed. To determine TR plasma properties, we fit a single-Gaussian (SG) model to the TR \ion{Si}{iv}~1394~\AA\ line. This was done using the procedure iris\_auto\_fit.pro available in the SolarSoft\footnote{https://www.lmsal.com/solarsoft} (SSW) library. In regions of the jet where strong deviations from a SG profile were detected, double-Gaussian (DG) fits were also performed using the SSW procedure dgf\_1lp.pro, which was modified slightly to suit our needs.

\begin{figure*}
\centering
\includegraphics[width=18cm]{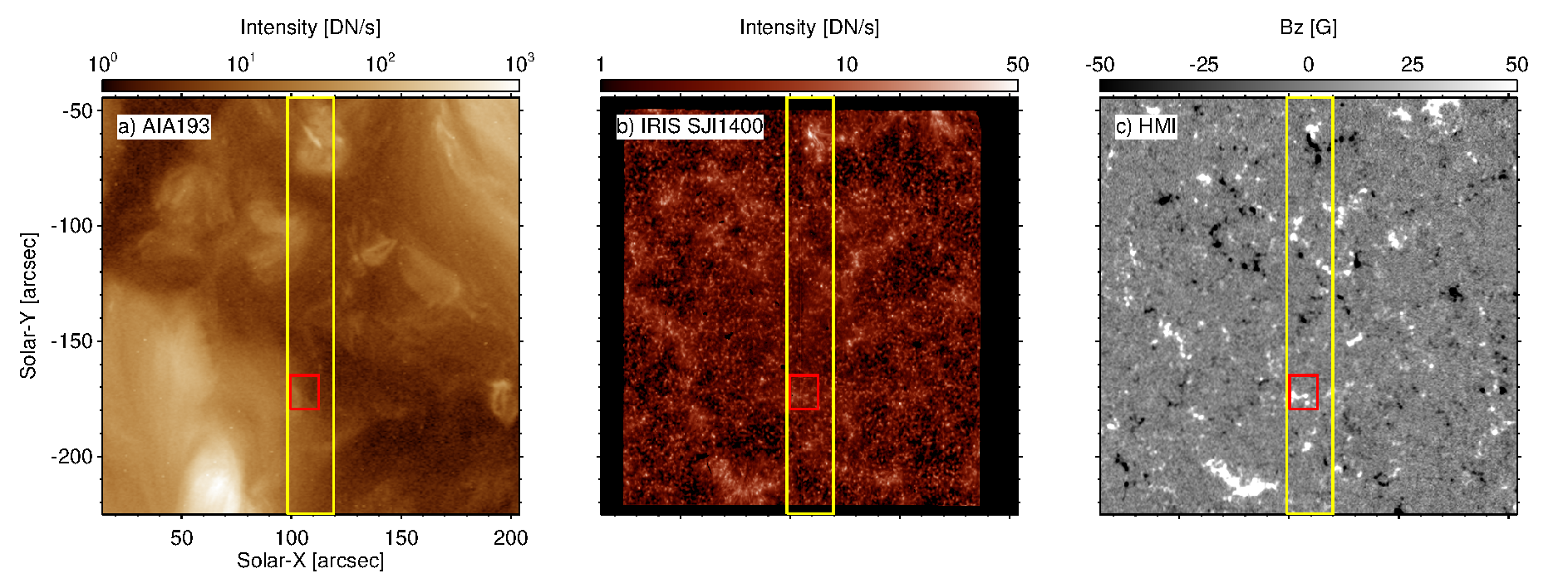}
\caption{Co-aligned maps of observation for AIA193, IRIS slit-jaw imager (SJI) 1400 channel, and HMI line of sight (LOS) magnetogram. Each image is selected at the nearest observation time to when the jet is seen in IRIS SJI1400 at 07:21~UT. The FOV for the IRIS raster scan is outlined in yellow and the red box marks the region of interest for the network jet, which is also the FOV for panel (b) in Fig.~\ref{fig:SJIs_B_dopp}. See Sect. \ref{section:Results}.}
\label{fig:AIA193_IR_HMI}
\end{figure*}

\section{Results}
\label{section:Results}
Overview imagery reveals the solar environment in which the jet originates. Figure~\ref{fig:AIA193overview} shows the full Sun as imaged by the 193~\AA\ channel of AIA at the time of the jet observation. A crescent-shaped CH is flanked on opposite sides by two bright ARs. The jet is located at the border between the CH and the lower AR. This general environment of the jet is more readily seen in Fig.~\ref{fig:AIA193_IR_HMI}, which is zoomed into the slit-jaw FOV. The jet emanates from a magnetic network region, as outlined by bright patches in IRIS SJI1400 intensity (panel b) and HMI LOS magnetic field concentrations (panel c).
\begin{figure}
\includegraphics[width=8.8cm]{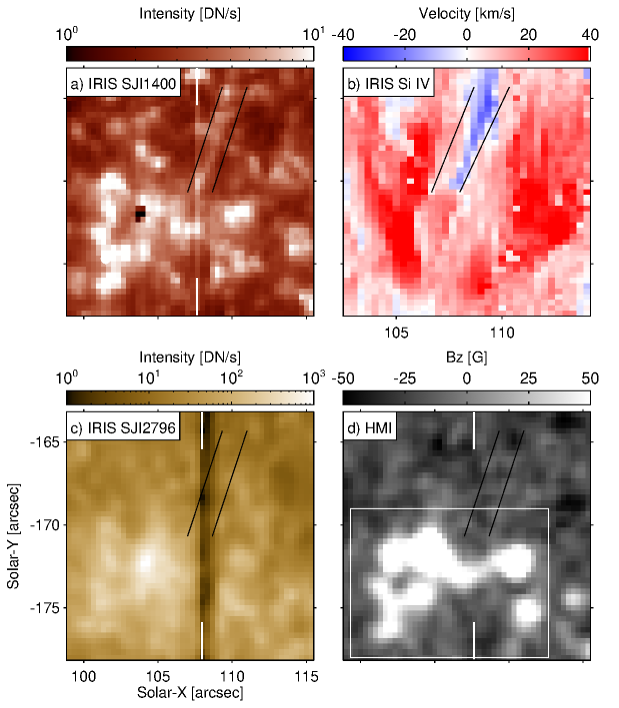}
\caption{Close-up view of jet. The region of interest for the network jet is shown from four different instruments: (a) IRIS SJI1400, (b) IRIS raster Dopplergram for \ion{Si}{iv} 1394~\AA, (c) IRIS SJI2796, (d) HMI LOS magnetogram. For (b), the FOV is the same as that demarcated in red in Fig.~\ref{fig:AIA193_IR_HMI}. However, for panels (a), (c), and (d), the FOV has been extended 3.3\arcsec\ (10 pixels) to the east (which is outside of the raster scan's spatial coverage) in order to fully encompass the observed activity at the jet's magnetic footpoint. For each panel, the jet is outlined in black. (a), (c), and (d) are co-aligned images taken at 07:21~UT and the location of the raster slit at that time is overplotted with white whiskers. The time evolution for panel (d) is available as an online movie. Panel (b) is calculated from single-Gaussian fits applied to the original IRIS raster data and has a smaller FOV due to the limited spatial extent of the scan. See Sect.~\ref{section:Results} for discussion.}
\label{fig:SJIs_B_dopp}
\end{figure}
\subsection{Jet in the transition region}
\label{subsection:Results_TR}
The jet's immediate surroundings across different layers of the solar atmosphere, as shown in Fig.~\ref{fig:SJIs_B_dopp}, provide insight into the dynamics at play for this event. The clearest signal from the jet is seen in the SG-fitted Dopplergram as a strip of upflowing material (blueshifted region in panel b). Based on these fits, we find that the observed jet has a line of sight velocity reaching $-30$~km\,s$^{-1}$. 

This strip appears in the intensity imagery taken by the SJI1400 channel (panel a), and it resembles the jets seen by \citet{tian:2014}. However, in our observation, the jet is only found in a single SJI frame. This is due to the SJI's low cadence of 2~min. The jet is also not easily detectable in any of the other SJI channels or in AIA. Below the jet, there is an area of increased activity in the chromospheric SJI2796 channel (panel c), and this region overlies clumps of positive polarity network magnetic field concentration seen in the HMI magnetogram (panel d).

Since the jet is only seen in one image frame, the jet's lifetime and apparent velocity within the plane-of-sky cannot be well constrained using slit-jaw images. With a cadence of 127~s, the maximum lifetime is estimated as the time between three frames, or 254~s. However, the minimum duration of observed upward-flowing material can also be constrained as the time it takes the raster to sweep across the jet. There are 7 slit positions with blueshifts recorded along the jet, with a cadence of 32~s between each position. The minimum duration of upflows is then 224~s. Together with the above estimates, this implies that the jet lifetime should be somewhere between 224~s and 254~s, that is on the order of about 4~min.
\begin{figure*}
\centering
\includegraphics[width=18cm]{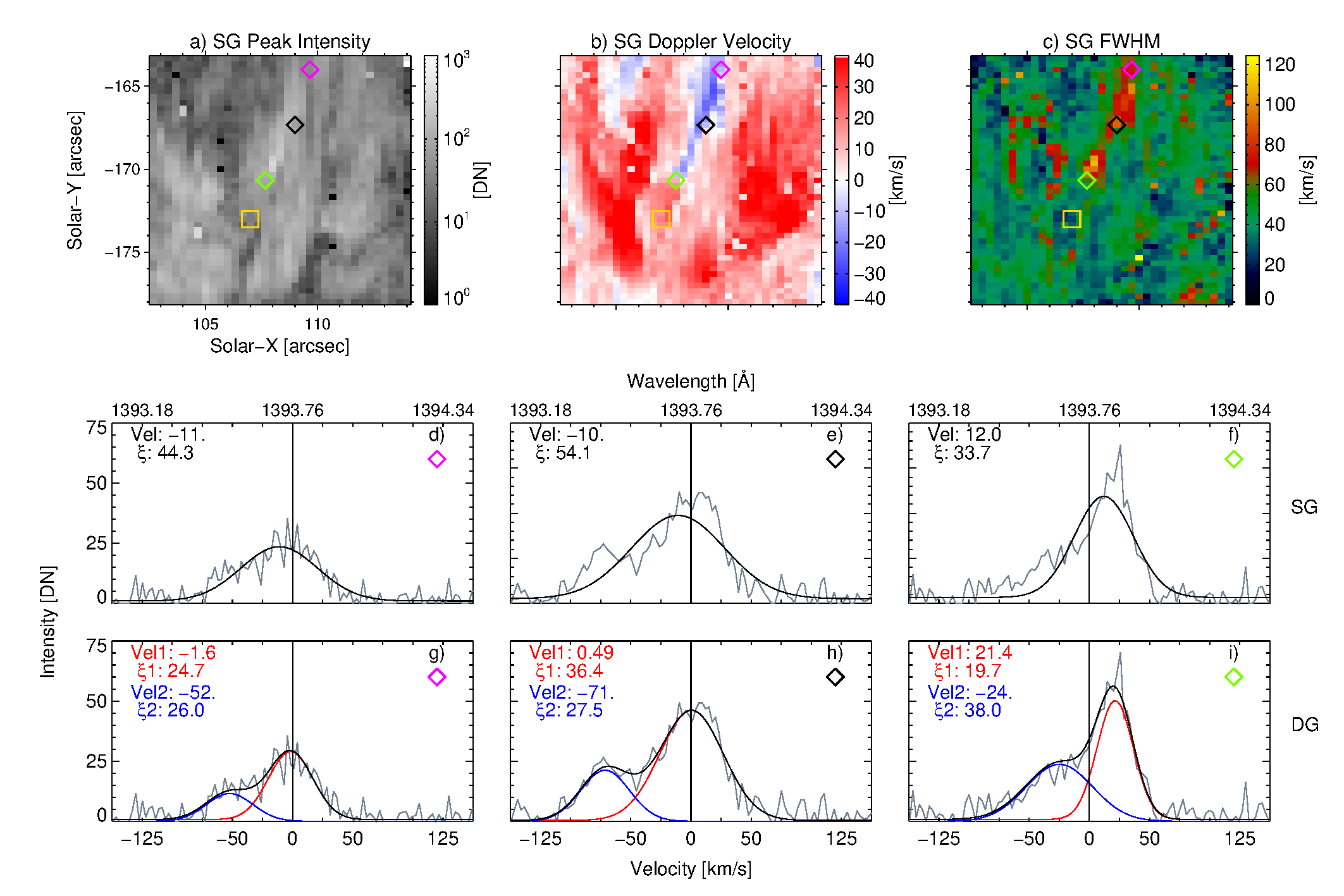}
\caption{\ion{Si}{iv} 1394~\AA\ Gaussian parameters and selected spectra from jet. Panels (a)--(c) show the single-Gaussian (SG) parameter maps for peak intensity [DN], Doppler velocity [km\,s$^{-1}$], and full-width at half maximum (FWHM) [km\,s$^{-1}$] for \ion{Si}{iv}, respectively. The colored diamonds mark the three individual pixels whose spectral profiles are shown in the bottom two rows, while the gold square marks, for reference, a fourth pixel whose spectrum is shown in Fig.~\ref{fig:IRspec_mg}. Panels (d)--(f) display the individual pixels' spectra in gray [DN] with the SG fitted curve overlaid in black. The Doppler velocity (Vel) and non-thermal width ($\xi$) are listed in the upper left. Panels (g)--(i) show the same three spectra, except a double-Gaussian (DG) fitted curve is overlaid in black and its two components are drawn in red (primary component) and blue (secondary component). The primary and secondary Doppler velocities (Vel1, Vel2) and non-thermal widths ($\xi 1$, $\xi 2$) are listed in the upper left. For all panels (d)--(i), a vertical black line in the center marks the location of zero-velocity, for reference. See Sects.~\ref{subsection:Results_TR} and \ref{subsection:Discussion_MultiComponents}.}
\label{fig:IRspec_dgsg}
\end{figure*}

Parameters derived from the SG fits of the \ion{Si}{iv}~1394~\AA\ line (formation temperature log\,$T$\,[K] = 4.8) outline the shape and extent of the jet within the transition region. Maps of the peak intensity, Doppler velocity, and full-width at half maximum (FWHM) derived from the SG fits are included in Fig.~\ref{fig:IRspec_dgsg}. They show that the jet consists of a narrow spire of increased intensity with broadened line widths. In comparison, directly beneath this spire are downflows (redshifted pixels) whose SG-fits have a lower peak intensity and FWHM. The Dopplergram shows a gradual increase in blueshift magnitude along the jet's spire from roughly 5--10~km\,s$^{-1}$ at its base up to 30~km\,s$^{-1}$ near the top.

While the parameter maps provide an overview of the event, the individual spectra are better at showing the flow structure within the jet. Spectra from three pixels representative of the upper, middle, and bottom portions of the jet are shown in panels (d)--(i) of Fig.~\ref{fig:IRspec_dgsg}, with either SG or DG fits applied. The SG fits plotted in the middle row seem to visually represent the observed intensities fairly well. However, most profiles along the jet show a two-component flow behavior that is best described by a DG fit (bottom row). This is supported by the reduced-chi-square metric, which is improved in virtually all cases where the DG fit is applied, with an average reduction by a factor of four. Further examples of DG-like intensity profiles are shown in the Appendix (Fig.~\ref{fig:Appendix1}). 

The spectral profiles along the length of the blueshifted segment of the jet (as seen in the SG Dopplergram) exhibit a secondary blueshifted component. Such profiles with a secondary blueshifted component are observed even at those locations that are redshifted when fit with a SG. At this base, the secondary component has a blueshift of $-13$~km\,s$^{-1}$, which increases to almost $-80$~km\,s$^{-1}$ farther up along the jet spire. The average secondary component velocity is $-51$~km\,s$^{-1}$ (see Figs.~\ref{fig:IRspec_dgsg} and \ref{fig:Appendix1}). When comparing blueshifts, Doppler velocities of the secondary component of the DG fit are always larger than those derived from the SG fits. Panels (e) and (h) of Fig.~\ref{fig:IRspec_dgsg} show the SG and DG curves for the same pixel in the middle of the jet. The DG fit secondary component has a velocity equal to $-71$~km\,s$^{-1}$, yet the SG fit velocity only has a line shift of $-10$~km\,s$^{-1}$. Comparable discrepancies can be seen all along the jet's spire. In addition to velocity, the secondary component has excess line widths ($\xi 2$; non-thermal velocity) that exceed typical non-thermal widths of 20~km\,s$^{-1}$ in the TR at $\approx$\,0.1~MK \citep{depontieu:2015}. We find that most of the secondary component non-thermal widths exceed this typical value, even by up to a factor of two or more. On average, the non-thermal width of the primary component is less than that for the secondary.

\begin{figure*}
\centering
\includegraphics[width=12cm]{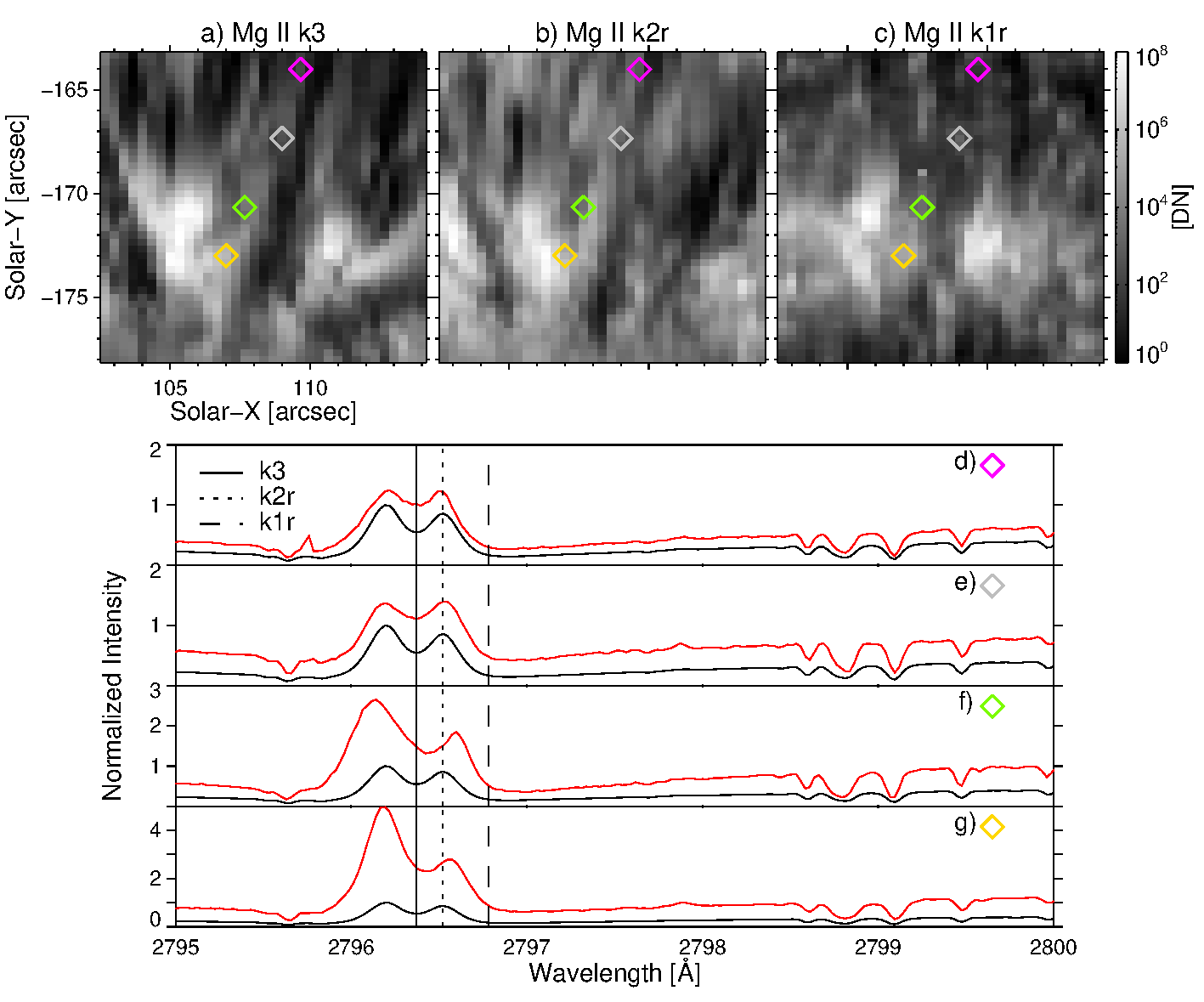}
\caption{Chromospheric activity at base of jet. Maps of the intensity [DN] at three wavelength points along the \ion{Mg}{ii} line are shown in the top row. The wavelength locations are: (a) k3: 2796.37~\AA, (b) k2r: 2796.52~\AA, and (c) k1r: 2796.78~\AA. Each map shares the same axis-scaling, spatial extent, and pixel markers as that in Fig.~\ref{fig:IRspec_dgsg}. \ion{Mg}{ii}~k spectra from these pixels are shown in the bottom four rows, panels (d)--(g). The black curve in each panel shows the same normalized, average intensity profile for \ion{Mg}{ii} for this observation, calculated using a large portion of quiet Sun. The k3, k2r, and k1r wavelength points were selected based on the shape of this profile and are marked by the vertical solid, dotted, and dashed lines, respectively. The k2v peak of this curve is used as the normalization factor for both the averaged profile and the individual pixel profiles (red curves). See Sect.~\ref{subsection:Results_Chromo}.}
\label{fig:IRspec_mg}
\end{figure*}

\subsection{Jet as seen in the chromosphere}
\label{subsection:Results_Chromo}
The jet's chromospheric activity is investigated using the \ion{Mg}{ii} 2796~\AA\ line, which is formed in the upper chromosphere within a temperature range of roughly log\,$T$\,[K]\,=\,3.7--4.2. The normalized \ion{Mg}{ii}~k intensity profiles from along the jet and at its base, as plotted in panels (d)--(g) in Fig.~\ref{fig:IRspec_mg}, show that, while the \ion{Mg}{ii} intensities along the upper spire of the jet are comparable to the QS profile, they become enhanced at its base (i.e., south of the region exhibiting blueshifts in \ion{Si}{iv}). These spectra located in the bright patch of activity beneath the jet can get to be around four times as intense as the QS (panel g). The intensity maps obtained around the k2r and k3 features of the \ion{Mg}{ii} line (panels (a) and (b) in Fig.~\ref{fig:IRspec_mg}) represent the intensities in the middle to top of the chromosphere at the base of transition region. There are still traces of the jet spire in these chromospheric layers. However, signatures of the spire are absent in the deeper layers of the chromosphere sampled with the \ion{Mg}{ii}~k1r feature (see panel c). Rather, only the bright patches (associated with the magnetic network) at the jet's footpoint are visible. 

It is possible to make inferences about the speed within the chromosphere by comparing the relative shifts and intensity variations along the \ion{Mg}{ii}~k line. Using the analysis procedure described in \citet{pereira:2013}, the pixels near the jet's footpoint show redshifts of the k3 depression that are consistent with downflows at the top of the chromosphere of about 5--10~km\,s$^{-1}$. Intensity asymmetries between the k2r and k2v peaks can also be exploited to reveal the presence of up- or downflows. When the k2v peak is enhanced compared to k2r, as is seen for this observation at the base of the jet, this signifies that plasma above the k2 formation height is moving downwards \citep[see][]{leenaarts:2013,pereira:2013}. The utility of chromospheric diagnostics using shifts of the k3 depression and comparison of the k2v to k2r peak intensities has also been analyzed and confirmed by \citet{bose:2019,bose:2021}. Thus, both of these diagnostics consistently and coherently reveal that there are downflows present in the upper chromosphere below or at the base of the jet.

\subsection{Magnetic structure at base}
\label{subsection:Results_MagBase}
The features detected in the LOS component of the magnetic field can be linked to the overlying jet structure. Comparing panels (c) and (d) from Fig.~\ref{fig:SJIs_B_dopp} shows that the increased chromospheric activity (i.e., radiation) beneath the jet is spatially aligned with strong concentrations of magnetic field. The raster maps from Fig.~\ref{fig:IRspec_mg} agree with this spatial overlapping of bright \ion{Mg}{ii} intensity and magnetic field patches. The time sequence of this region reveals that the resolved clumps of positive polarity magnetic field rearrange themselves, separating and coalescing, in the period of time surrounding the jet (see online movie associated with Fig.~\ref{fig:SJIs_B_dopp}d). 
\begin{figure*}
\centering
\includegraphics[width=12cm]{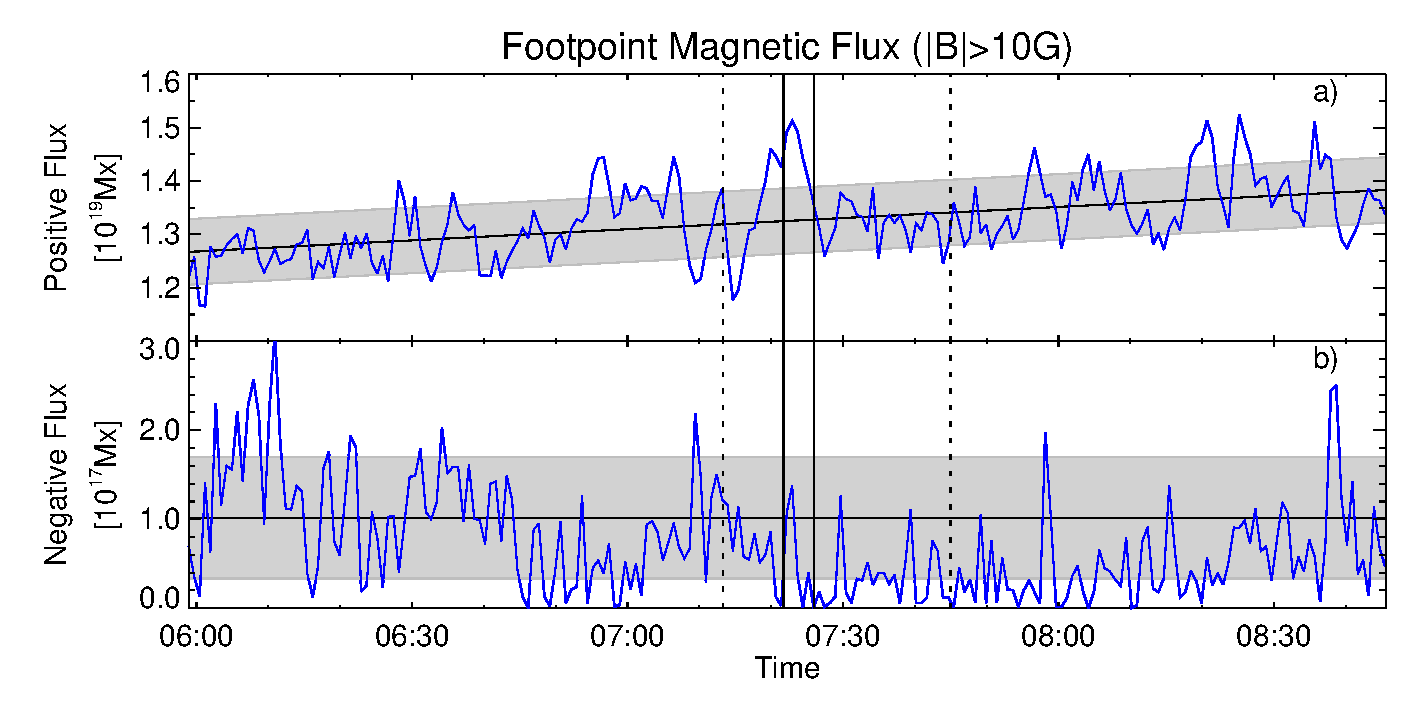}
\caption{Magnetic flux vs. time at jet's footpoint. Magnetic flux of positive (panel a) and negative (panel b) polarity magnetic concentrations within the footpoint region of the jet (white box in Fig.~\ref{fig:SJIs_B_dopp}d) for pixels stronger than 10~G are shown. The dotted vertical lines mark the start and stop times for the entire observation period of the IRIS slit-jaws: 07:13:15--07:44:58~UT. The solid vertical lines mark the period when the raster is scanning across the jet: 07:21:43--07:25:46~UT. The SJI1400 frame wherein the jet is seen occurs at 07:21:43~UT. In panel (a), the black line shows the linear trend. The gray shading covers the root-mean-square error (RMSE) from this fit. Panel (b) shows the root-mean-square (RMS) and the RMSE for the calculated values from the RMS by the black horizontal line and the gray shading, respectively. See Sects.~\ref{subsection:Results_MagBase} and \ref{subsection:Discussion_Driver}.}
\label{fig:MagFlux}
\end{figure*}

A more quantitative view of the magnetic evolution is provided by the time evolution of the magnetic flux. Figure~\ref{fig:MagFlux} illustrates how the magnetic flux, of both positive and negative polarity patches, at the footpoint of the jet changes with time. Here, the footpoint is defined by the white box in panel (d) of Fig.~\ref{fig:SJIs_B_dopp}. To avoid noise, only pixels with magnetic flux density greater than 10~G were integrated to derive magnetic flux (this was done separately for positive and negative polarity regions). Although, selecting different cut-off strengths of 5~G, 10~G, 15~G, or 20~G showed no significant difference in the time series of magnetic flux, underlining that our results are not affected by noise. Magnetic flux contained by negative polarity patches in this region is two orders of magnitude weaker than that of the positive polarity patches. In addition, variations in the magnetic flux of negative polarity magnetic concentrations are generally within the computed root-mean-square-error (RMSE) for the whole time series. For the magnetic flux of positive polarity fields, however, a few things are noted. First, a general, linearly increasing trend is detected over the observing period of three hours. The fit to this trend was used to calculate the RMSE. Around the time of the jet, the positive flux shows strong fluctuations above the level of the RMSE. In the ten minutes preceding the jet, magnetic flux rises sharply and then decreases significantly after the jet is seen in SJI1400. There are a few other instances of such strong changes in positive flux outside of the IRIS observing time, though none are of the same net magnitude seen during the jet's lifetime.  

\section{Discussion}
\label{section:Discussion}
Most TR jet studies have been conducted using slit-jaw imagery or sit-and-stare rasters. As a result of this, any actual mass motions along a jet's trajectory are not easily identified. Propagating intensity fronts in SJIs could correspond to moving heat fronts or wave motions. Stationary-slit rasters, when at the limb, would only likely pick up transverse motions. When near disk-center, such rasters typically only cross small sections of a jet, thereby making it difficult to determine if upward flows are persistent along the rest of the structure. With this observation, however, we have the good luck that the raster scan was both co-temporal and co-spatial with the jet motions along its length, allowing for verification of the presence of mass flows and deeper inspection of spectral profiles. 

\subsection{Jet properties}
\label{subsection:Discussion_Properties}
Various jet properties, including the propagation speed and lifetime, can be estimated but not fully constrained, owing to various limitations of the observation. Due to the unknown inclination of the jet, exact speeds along its trajectory cannot be precisely calculated. Using the plane-of-sky propagation of the jet based only on the spatial extent of the blueshifted pixels as seen in Figs.~\ref{fig:SJIs_B_dopp}b and \ref{fig:IRspec_dgsg}b ($\approx$~20 pixels amounting to a total length of 5~Mm) and the time it takes the raster to scan across them (7 scans at a cadence of 32~s amounting to 224~s total), it can be estimated as a lower bound that the jet propagates laterally at 22~km\,s$^{-1}$. If we further assume a simplistic triangular geometry, where the plane-of-sky speed calculated above is used for the horizontal speed and the average LOS velocity of the jet material represents the vertical velocity (51~km\,s$^{-1}$, see Sect.~\ref{subsection:Results_TR}), then we can calculate a propagation speed of 55~km\,s$^{-1}$. This speed is 1.5--4.5 times lower than that determined for typical CH jets by \citet{tian:2014} and \citet{narang:2016}, but that leads us to presume the jet is traveling with speeds greater than the observational cadences are able to resolve. The projected length of the blueshifted material does agree with previous estimates \citep{tian:2014,narang:2016}, rounding out to be between 4.5 and 5.2~Mm. As we stated previously, with only one slit-jaw image showing the jet, it is impossible to provide accurate estimates of the jet's lifetime. An upward estimate of 4~min~14~s, or the time between three slit-jaw images, grossly outlasts the average established jet lifetime of 20--80~s, as given by \citet{tian:2014}. This number can be further winnowed, albeit slightly, by considering how long the raster takes to scan across the jet. In that case, a lifetime of 3.75~min is calculated (see Sect~\ref{subsection:Results_TR}). 

\subsection{Multiple flow components}
\label{subsection:Discussion_MultiComponents}
Plasma outflows (blueshifted region along the spire) confirm the presence of real mass motions. The physical extent of the jet actually extends beyond what is inferred solely from SG blueshifts of \ion{Si}{iv} (see e.g., Fig.~\ref{fig:IRspec_dgsg}). The jet is traced down to its region of origin by analyzing the individual spectra, whose intensities reveal asymmetric profiles. These asymmetries are well-described by double-Gaussian fits to the observed \ion{Si}{iv} line profiles. The primary component of the DG is representative of the typical TR redshifted material emitting beneath (i.e., below) the source region of the secondary component, which corresponds to the blueshifted jet material. By evaluating the secondary component, the behavior of the jet is elucidated.

Previous work has investigated the common appearance of asymmetrical emission lines formed in the corona and TR \citep[see e.g.,][]{peter:2001,hara:2008,tian:2011,yardley:2021}. As is the conclusion of this work, such line asymmetries are mostly attributed to the overlapping of background emission (primary component) and high-speed flows (secondary component). \citet{tian:2011} discovered a link between the secondary emission component and upward propagating disturbances seen in AIA. Ubiquitous, propagating torsional motions associated with dynamic features within the TR and chromosphere were first observed at small scales with IRIS and have been reported by \citet{depontieu:2014_10}. \citet{kayshap:2018} attribute a change in Doppler shift of the secondary component from red to blue across network jets to twisting motions. For the event we describe here, however, the DG profiles are not interpreted as signatures of large-scale twist. One reason for this is that the secondary component remains blueshifted compared to the primary component, regardless of which side of the feature we regard. While there may be unresolved plasma motions that can be interpreted as small-scale torsional motions within the jet as evidenced by the non-thermal widths, a general, coherent overturning rotation of the structure is not supported by the observational data.

The blueward flow component associated with the jet material is fast enough that it might contribute to the solar wind if the magnetic fields that confine the jet open into the heliosphere. This secondary component exhibits vertical speeds up to 50--75~km\,s$^{-1}$ (see Fig.~\ref{fig:IRspec_dgsg}). Comparing this to the sound speed in the source region of \ion{Si}{iv} that forms in ionization equilibrium at around 80\,000~K ($\approx$\,43~km\,s$^{-1}$), these outward flows are supersonic. Such high-speed flows accelerate further as they ascend into progressively less dense plasma, and, if connected to an open field line within the CH, could escape into the solar wind. However, in order to be comparable with the solar wind speeds of a few hundred km\,s$^{-1}$ recorded at 1~AU, these TR velocities would need to accelerate by at least a factor of six or more (and then the plasma has to be lifted out of the gravity field of the Sun). Hence, further acceleration would be required to eject this plasma as solar wind. 

An extra source of energy to propel the jet into the solar wind could be derived from its non-thermal velocities. The jet has non-thermal widths of up to 45~km\,s$^{-1}$ along its spire (see Fig.~\ref{fig:IRspec_dgsg}). If these unresolved motions are assumed to be wavelike in nature, caused by an Alfv\'en wave, an energy flux can be calculated and compared to what is needed in terms of extra energy. The Alfv\'en wave energy flux, $f$, can be estimated from:
    \begin{equation}
        f = \rho\, \xi^2~ V_A \,,
    \end{equation}

where $\rho$ is the mass density, $\xi$ is the non-thermal velocity, and $V_A$ is the Alfv\'en speed. We use the average calculated non-thermal velocity of 34~km\,s$^{-1}$ and the propagation speed estimated in Sect.~\ref{subsection:Discussion_Properties}, as well as a mass density of $1.8 \times 10^{-14}$~g\,cm$^{-3}$, which is calculated assuming a typical TR density of $10^{10}$~cm$^{-3}$ \citep[see Supplementary Materials in][]{tian:2014}. This results in an energy flux of 1.2~kW\,m$^{-2}$, which is an order of magnitude greater than the 100--200~W\,m$^{-2}$ needed on average to drive the solar wind \citep{schwenn:1990}. Certainly, this is a significant amount of energy that could travel upward and boost the jet material at higher altitudes, thereby powering the solar wind. However, it could also be the case that the energy is not carried up high enough and is dissipated before it reaches a regime favorable for acceleration into the wind.

\subsection{Magnetic driver}
\label{subsection:Discussion_Driver}
We can now look towards determining the jet's driver. We see signatures of increased activity in the \ion{Mg}{ii} line extending below the jet's line of expression in \ion{Si}{iv}. This intense patch is the presumed region of origin for the jet as it is located above strong concentrations of vertical magnetic field. 

Unfortunately, the behavior of the magnetic field at the jet's footpoint on both long and short timescales does not point unambiguously to a specific driving mechanism. Overall, no obvious signs of systematic flux cancelation are detected, but the positive flux linearly increases by around 9\% across a period of three hours, which is potentially a sign of flux emergence (see Sect.~\ref{subsection:Results_MagBase} and Fig.~\ref{fig:MagFlux}). Interesting fluctuations, which are significant above the threshold set by the RMSE-level, are seen over the timescale during which the jet evolves. The sharp increase in positive magnetic flux immediately preceding the jet followed by a stark decrease during the jet's activity might be a signature of local reconnection (see the evolution in Fig.~\ref{fig:MagFlux}a from 07:20--07:25~UT indicated by the vertical lines). Similar magnetic field fluctuations before and after the event, while not quite as intense, could also be potential signatures of persistent jet activity in this region. However, a lack of IRIS coverage at those times makes it impossible to say whether or not this is true.

Though large-scale flux cancelation is not obvious from the HMI data due to a lack of noticeable available negative field concentrations, the movie of the magnetic field at the base of the jet shows merging and separation of positive magnetic clumps (see online movie associated with Fig.~\ref{fig:SJIs_B_dopp}d). This could further support the theory of small-scale reconnection between the strong, resolved positive magnetic flux and nearby, unresolved areas of negative flux. The plausibility of such a scenario is demonstrated by \citet{chitta:2017:sunrise} who show that, using the balloon-borne mission SUNRISE \citep{solanki:2010,solanki:2017}, small jets at the base of coronal loops are driven by flux cancelation and reconnection between large-scale, dominant polarity patches and small-scale elements of the opposite polarity. When comparing the same observational area using HMI, however, the non-dominant polarity is not clearly seen. Similarly, \citet{chitta:2019} observe transient magnetic flux emergence and cancelation events using the Swedish 1-m Solar Telescope \citep[SST;][]{scharmer:2003}, which has five times the resolution as HMI. These transient magnetic events are seen at granular spatial scales and are found to appear in photospheric plage regions with a predominantly unipolar vertical magnetic field. The authors compare these observations with 3D radiation magnetohydrodynamic (rMHD) simulations, verifying the presence of these magnetic events wherein emerging, small-scale and non-dominant polarity magnetic elements cancel with the surrounding dominant unipolar magnetic field. They find that these events can appear on even smaller spatial and temporal scales than that resolved by observations. 

\citet{chen:2019} discuss the relationship between explosive events (EEs) and network jets. Two relevant categories of EEs that the authors define are for line profiles displaying either two peaks of comparable magnitude or enhancement at the blue wing, and they label these as categories (ii) and (iii), respectively. They find that EEs of the latter group are mainly located on network jets, away from the jet footpoints, while those of the former are either seen at the footpoints of jets or on transient compact brightenings not associated with network jets. For the jet evaluated in our study, the spectral profiles along the jet's spire show strong blue wing enhancements similar to those of category (iii), while closer to the base, some profiles exhibit category (ii)-type double peaks of commensurate intensity (see e.g., Appendix Fig.~\ref{fig:Appendix1}, panel k). We find here that not only are the locations of the different profile types consistent with the previous work, but they also further the proposal of reconnection stemming from the shifting magnetic elements identified at the footpoint. 

\subsection{Location of energy release}
\label{subsection:Discussion_EnergyReleaseLocation}
In our observation, we find that the chromospheric plasma mainly exhibits downflows at the jet's base, whereas plasma sampling the TR at 0.1~MK is propelled upwards and thus blueshifted (see Sects.~\ref{subsection:Results_TR} and \ref{subsection:Results_Chromo}). This suggests that the actual location of energy release is located between the up- and downflows; in response to the energy deposition, the gas is propelled away from the release site and is channeled by the magnetic field in roughly up- and downward directions. 

This is essentially the same setup as in an EE \citep{dere:1989}, where spectroscopic observations showed a bi-directional outflow from a reconnection site \citep{innes:1997}. In those EEs, the up- and downflows were seen in the same spectral line, meaning they happened at the same temperature. This might be due to the fact that the reconnection process is happening in a low-beta regime (well above the chromosphere) where the plasma is locally heated by plasmoid-mediated reconnection \citep{peter:2019}. Observations suggest that in (some) EUV bursts the X-type neutral point where the reconnection occurs is located within the chromosphere some 500\,km above the photosphere, closer to where plasma beta is near unity \citep{chitta:2017}. Still, if the energy deposition is strong enough, those EUV bursts could show up- and downflows in a TR line like \ion{Si}{iv} \citep{peter:2014}, even if the reconnection is located almost as deep in the atmosphere as it is for Ellerman bombs \citep{georgoulis:2002}.

Our scenario that these network jets are driven from the top of the chromosphere is similar to earlier concepts, albeit in a different context. In previous works, a dichotomy of red- and blueshifts is seen in the QS TR (red) and corona (blue) in terms of temperature -- below about 0.5~MK the plasma is redshifted, while above 0.5~MK it is blueshifted \citep{peter:1999_05,peter:1999_09}. This can be understood in terms of the 3D MHD model of \citet{hansteen:2010}. In their work, the authors showed that the highest energy deposition per particle was found in the middle TR through episodic heating events. In response, the plasma expands and produces the redshifts of the TR and blueshifts of the coronal lines. 

Applying this model to our jet scenario, where energy is deposited at the top of the chromosphere, localized plasma heating creates a plug of increased pressure. The gas expands along field lines on both sides, which leads to downflows into the upper chromosphere seen in \ion{Mg}{ii} and upflows towards the corona seen in \ion{Si}{iv}. While the ultimate driver might be found in the photospheric stretching of the magnetic field \citep[e.g.,][]{priest:2002}, the actual energy release that drives the jet we observe might be released at the top of the chromosphere.

\section{Conclusion}
\label{section:Conclusion}
We have observed a network jet with mass motions as inferred from secondary blueshifted components (20--70~km\,s$^{-1}$) in the spectral profiles. By comparing the magnetic field data with spectral information from the \ion{Mg}{ii} and \ion{Si}{iv} lines, the jet has been followed from its footpoint location in the photosphere up through the chromosphere and into the TR. 

We find that the energy that drives the network jet is released at the top of the chromosphere such that it drives parts of the plasma downwards into the chromosphere and other parts upwards into the upper atmosphere. The line broadening suggests that there should be sufficient surplus energy available to drive the jet plasma to join the solar wind originating from the coronal hole wherein this jet is embedded.

This work is serendipitous in that the raster slit happened to be scanning across roughly the length of a jet just as it was accelerating. However, the findings are limited to a single jet that is only seen in one SJI frame. Thus, future work should attempt to obtain similar observations that round out these shortcomings. Such similar observations would be dense raster scans located within or at the boundary of CHs at disk center with accompanying slit-jaw imagery. Preferably, the slit-jaw imagery would have a higher cadence than the observation presented herein (127~s) in order to capture any jet events in more than one frame; however, the cadence should not be too high so as to degrade the S/N beyond the point of utility. In addition to having multiple, similarly well-aligned raster observations of small-scale jets, higher resolution magnetic field data would help to better constrain possible driving mechanisms. Also, higher cadence SJIs would be useful for directly comparing the jet motions in the image plane with its motions along the LOS.

\section{Acknowledgments}
We thank the anonymous referee for providing helpful comments and suggestions. This work was supported by the International Max-Planck Research School (IMPRS) for Solar System Science at the University of Göttingen. IRIS is a NASA small explorer mission developed and operated by LMSAL with mission operations executed at NASA Ames Research Center and major contributions to downlink communications funded by ESA and the Norwegian Space Centre. SDO/AIA and HMI imagery and data are courtesy of NASA/SDO and the AIA, EVE, and HMI science teams.


\onecolumn
\begin{appendix}
\section{Further spectra examples from jet}
Here, we include more examples of double-Gaussian fitted spectra from along the jet. These are discussed in Sect. \ref{subsection:Discussion_MultiComponents}.
\begin{figure*}[h]
\centering
\includegraphics[width=18cm]{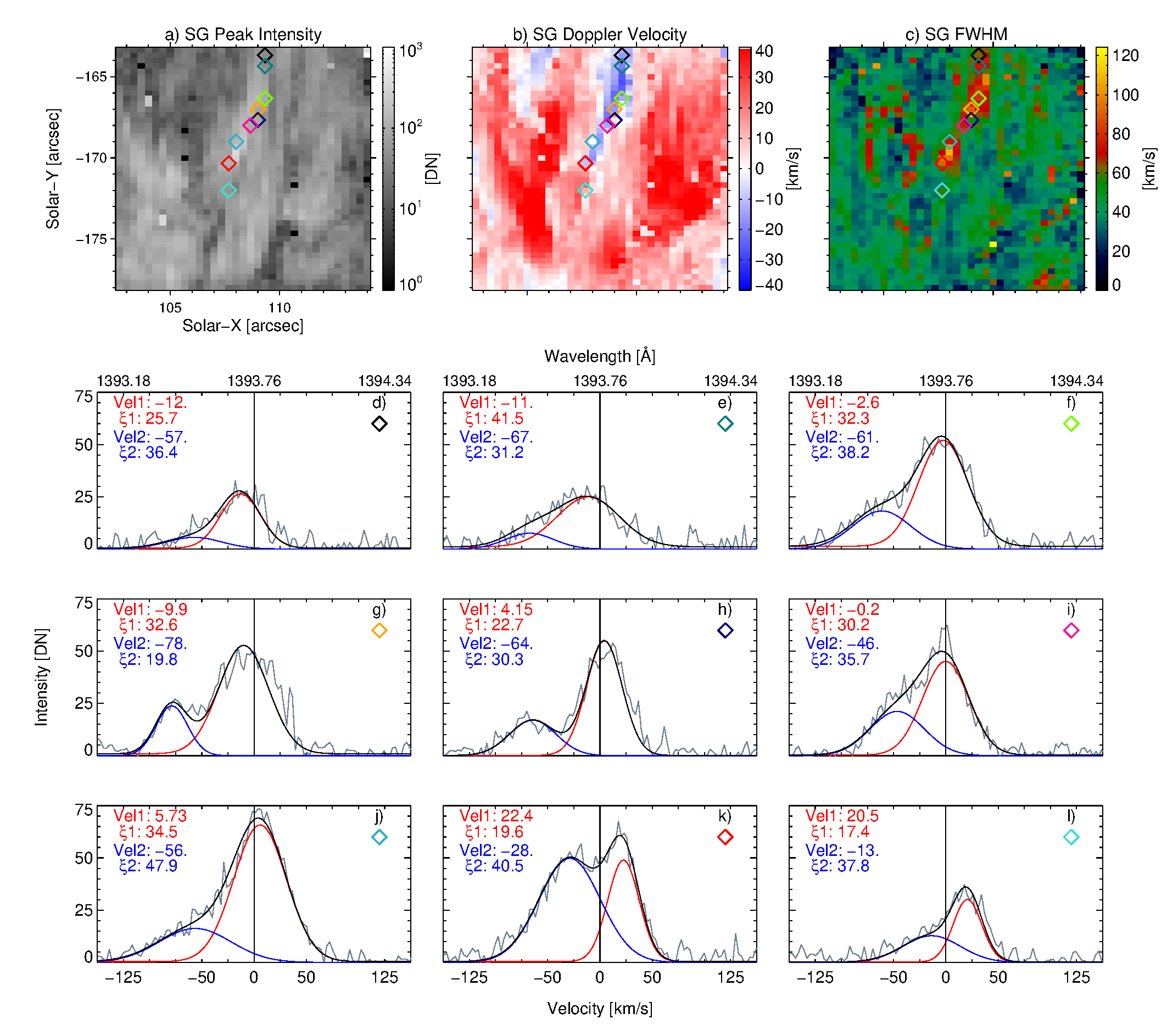}
\caption{More DG-fitted spectra along jet. Panels (a)--(c) are the same as in Fig.~\ref{fig:IRspec_dgsg}, except for the colored diamond markers, which mark nine different pixels whose spectral profiles are shown in the bottom three rows. The style and layout of the spectra in panels (d)--(l) also match those for panels (g)--(i) in Fig.~\ref{fig:IRspec_dgsg}.}
\label{fig:Appendix1}
\end{figure*}

\end{appendix}

\end{document}